\def\target{IC}
\def\CACM{CACM}
\def\CCR{CCR}
\def\IC{IC}

\ifx\target\CCR
\documentclass[9pt,sigconf,screen=true]{acmart}

\setcopyright{none}
\renewcommand\footnotetextcopyrightpermission[1]{}

\usepackage{fancyhdr}
\fancypagestyle{plain}{%
   \fancyhf{} %
   \fancyfoot[L]{ACM SIGCOMM Computer Communication Review}%
   \fancyfoot[R]{Volume 51 Issue 1, January 2021}%
}
\fancypagestyle{firstpagestyle}{%
   \fancyhf{} %
   \fancyfoot[L]{ACM SIGCOMM Computer Communication Review}%
   \fancyfoot[R]{Volume 51 Issue 1, January 2021}%
}
\fi

\ifx\target\CACM
\documentclass[sigconf,screen=true]{acmart}
\setcopyright{none}
\renewcommand\footnotetextcopyrightpermission[1]{}

\acmConference[CACM '21]{Communications of the ACM}{MM 2021}{Vol. X No. Y}
\acmYear{2021}
\copyrightyear{2021}

\fi

\ifx\target\IC
\documentclass{IEEEcsmag}

\usepackage[hyphens]{url}
\usepackage[colorlinks,urlcolor=blue,linkcolor=blue,citecolor=blue]{hyperref}

\usepackage{upmath}

\jvol{XX}
\jnum{XX}
\paper{XX}
\jmonth{November/December}
\jname{Internet Computing}
\pubyear{2022}

\fi

\usepackage[utf8]{inputenc}
\usepackage[T1]{fontenc}
\usepackage[nolist]{acronym}
\usepackage{url}
\graphicspath{{figures/}}

\usepackage{pifont}
\usepackage{multirow}
\usepackage{siunitx}
\usepackage{array,booktabs}

\usepackage[caption=false]{subfig}

\usepackage{color,soul}

\definecolor{gray}{gray}{0.9}
\definecolor{white}{gray}{1.0}
\newcolumntype{g}{>{\columncolor{gray}}l}
\newcolumntype{x}{>{\columncolor{gray}}p{1cm}}
\newcolumntype{y}{>{\columncolor{gray}}p{4cm}}
\newcolumntype{z}{>{\columncolor{gray}}p{2cm}}

\usepackage{todonotes}

\begin{acronym}
    \acro{CDN}{Content Delivery Network}
    \acro{AS}{Autonomous System}
    \acro{QoE}{Quality of Experience}
    \acro{QoS}{Quality of Service}
    \acro{CNAME}{canonical name}
    \acro{IPFS}{InterPlanetary File System}
    \acro{IPNS}{InterPlanetary Name System}
    \acro{IPLD}{InterPlanetary Linked Data}
    \acro{P2P}{peer-to-peer}
    \acro{ICN}{Information-Centric Networking}
    \acro{DHT}{Distributed Hash Table}
    \acro{URL}{Uniform Resource Locator}
    \acro{SFS}{Self-certifying File System}
    \acro{DAG}{Directed Acyclic Graph}
    \acro{GDPR}{General Data Protection Regulation}
    \acro{PKI}{Public-Key Infrastructure}
    \acro{CA}{Certificate Authority}
    \acro{CID}{Content Identifier}
\end{acronym}

\acrodefplural{AS}[ASes]{Autonomous Systems}
\acrodefplural{CDN}[CDNs]{Content Delivery Networks}
\acrodefplural{DHT}[DHTs]{Distributed Hash Tables}
\acrodefplural{URL}[URLs]{Uniform Resource Locators}
\acrodefplural{PKI}[PKIs]{Public-Key Infrastructures}
\acrodefplural{CA}[CAs]{Certificate Authorities}
\acrodefplural{CID}[CIDs]{Content Identifiers}

\ifx\target\CCR
\begin{teaserfigure}
	\parbox{\textwidth}{\centering\normalsize
		This article is an editorial note submitted to CCR. It has NOT been peer reviewed.\\
		The authors take full responsibility for this article's
		technical content. Comments can be posted through CCR Online.
	}
	\vspace{10pt}
\end{teaserfigure}
\fi

\begin{document}


\title{Towards Decentralised Cloud Storage with IPFS: Opportunities, Challenges, and Future Considerations}

\ifx\target\IC
\author{Trinh Viet Doan}
\affil{Technical University of Munich}

\author{Yiannis Psaras}
\affil{Protocol Labs}

\author{Jörg Ott}
\affil{Technical University of Munich}

\author{Vaibhav Bajpai}
\affil{CISPA Helmholtz Center for Information Security}

\markboth{}{Paper title}

\else
\author{Trinh Viet Doan}
\affiliation{
    \institution{Technical University of Munich}
    \country{{}}
}
\email{doan@in.tum.de}

\author{Yiannis Psaras}
\affiliation{
    \institution{Protocol Labs}
    \country{{}}
}
\email{yiannis@protocol.ai}

\author{Jörg Ott}
\affiliation{
    \institution{Technical University of Munich}
    \country{{}}
}
\email{ott@in.tum.de}

\author{Vaibhav Bajpai}
\affiliation{
    \institution{CISPA Helmholtz Center for Information Security}
    \country{{}}
}
\email{bajpai@cispa.de}
\fi

\begin{abstract} 
    %
    %
    The \acf{IPFS} is a novel decentralised storage architecture, which attempts to provide
    decentralised cloud storage by building on founding principles of \acs{P2P} networking and
    content addressing. 
    IPFS is used by more than 230k peers per week and serves tens of millions of requests per day, which makes it an interesting large-scale operational network to study. While it is used as a building block in several projects and studies, its inner workings, properties, and implications have only been marginally explored in research.
    Thus, we provide an overview of the IPFS design and its core features, along with
    the opportunities that it opens as well as the challenges that it faces because of its properties. Overall, IPFS presents an interesting set of characteristics and offers lessons which can help building decentralised systems of the future.
\end{abstract}

\maketitle

\ifx\target\IC
\else
\pagestyle{empty}
\fi



\section{Introduction}
\label{sec:introduction}

\ifx\target\IC
\chapterinitial{The prevailing baseline Internet infrastructure}
\else
The prevailing baseline Internet infrastructure
\fi
is based on centralised cloud storage and management, as large cloud and \ac{CDN} providers are seen to store significant amounts of user data in data silos.
%
Simultaneously, these providers control most of today's Internet traffic, which results in a substantial centralisation of data and traffic.
To counteract this, several initiatives and groups have focused on developing decentralised alternatives. However, the decentralised architecture of such networks may also add new challenges and types of complexity.  In this paper, we discuss these issues --- both the benefits and challenges --- in relation to InterPlanetary File System (IPFS)~\cite{Benet14}, a protocol for decentralised cloud storage.


IPFS is an open-source set of protocols that combines multiple existing concepts from \ac{P2P} networking, Linked Data, and other areas to allow participants to exchange
pieces of files, similar to Bittorrent. To simplify the retrieval of files, content on IPFS is
uniquely named and addressed using the so called \emph{multihash}, which is a self-describing datatype that adopts concepts from git's versioning model, cryptographic hashing, and Merkle Trees. In this resulting naming scheme, content is identified and accessed using names, rather than through location-based
identifiers such as \acp{URL}, as is also the case with several \ac{ICN} architectures~\cite{XylomenosVSFTVKP14}. In practice, IPFS integrates important components from several projects to enable
content distribution and availability in a decentralised manner.
For instance, the IPFS network hosts snapshots of Wikipedia to circumvent  censorship of information~\cite{WikipediaIPFS}.

\textbf{Deployment figures as of 2022.}
The IPFS network has been gaining constant momentum over the last years: The \texttt{ipfs.io} public gateway (hosted by \emph{Protocol Labs}, the primary maintainer of the open-source IPFS project) sees 3.7M unique users and serves more than 125TBs of data in more than 805M requests per week as of 2022~\cite{ipfs-2021}.
Such IPFS gateways act as web-servers for users outside the IPFS network (i.e., users
that do not participate as peers themselves) and carry out requests on behalf of those users (see~\autoref{sub:gateway} for more details).
IPFS network measurements~\cite{Henningsen20} in 2020 report numbers of roughly 6k publicly reachable nodes on average, with a much larger number of nodes not being reachable due to NAT. The authors' continued periodic measurements\footnote{\url{https://trudi.weizenbaum-institut.de/ipfs_analysis.html}} since then show that the number has increased to roughly 17k reachable nodes per crawl as of 2022, suggesting a substantial growth over the last few years. Overall, \emph{Protocol Labs} estimates the number of distinct active nodes per week to be more than 230k nodes, which makes the IPFS network one of the largest permissionless and decentralised P2P storage and delivery networks in operation.

As such, it has also found significant support among several Web-related projects with the goal of future-proofing their products:
For instance, Cloudflare initially started hosting IPFS gateways in more than 150 of
their data centers in September 2018,
(later increasing to more than 200), which are still operational and serve the IPFS network to date.
Mozilla Firefox added the \texttt{ipfs://} scheme to the list of allowed custom protocols in March 2018~\cite{FirefoxSupport} to support protocol handlers for browser extensions. As of January 2021, the Brave browser added native support for IPFS~\cite{brave-ipfs}, making it possible to access \texttt{ipfs://} links directly from the browser window, similar to the native IPFS support in Opera browsers~\cite{ipfs-2021} since March 2020.

Previous work
primarily studied IPFS as a storage mechanism for
specific use cases, such as IoT and edge computing~\cite{DBLP:conf/iot/AliDA17,DBLP:conf/icfec/ConfaisLP17,DBLP:conf/esocc/KrejciS020}, malware~\cite{DBLP:journals/ijisec/PatsakisC19,DBLP:journals/access/CasinoPAP20}, or blockchain technology~\cite{DBLP:conf/webi/ZhengLCD18,DBLP:conf/ithings/NorvillPSC18}. In particular,
IPFS already plays a significant role in paving the way for decentralised applications as the reference \ac{P2P} storage solution for hundreds of projects. However, the
socio-technical impact of decentralised architectures such as IPFS have not been explicitly weighed yet, although IPFS and similar P2P-based data storage architectures are already used in several proposals and services~\cite{Daniel2022}.
Towards this end, we first provide an overview of the design and building blocks of IPFS
(\autoref{sec:building-blocks}) in this paper. We discuss its properties along with associated socio-technical opportunities and
challenges (\autoref{sec:properties}),
before highlighting open questions that warrant further research (\autoref{sec:future-directions}), all of which can also be applied to build and improve other decentralised systems.

Note that the goal of this paper is to describe and discuss technicalities of IPFS together its
opportunities and challenges in order to distill lessons learned and open research questions for
future studies. We further do not focus on legal or economic aspects of IPFS; while we do discuss
legal issues, we do so in the context of its technical properties.

\section{Building Blocks and Principles}
\label{sec:building-blocks}

\begin{figure*}[t]
    \centering
    \includegraphics[width=1.5\columnwidth]{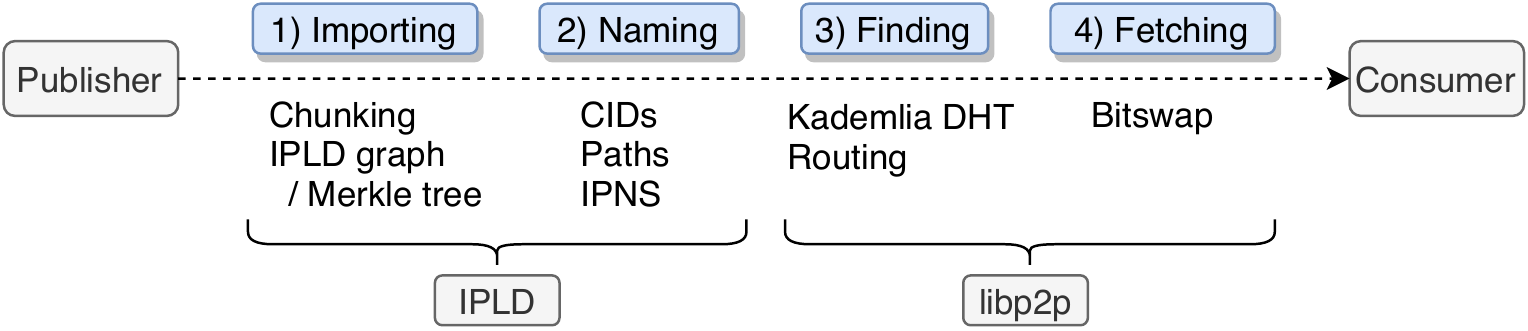}
    \caption{Outline of content publication (importing and naming) and retrieval process (finding and fetching), along with the involved IPLD and libp2p protocol stacks.}
    \label{fig:content-pub-retr}
\end{figure*}

The design of IPFS, originally described in its whitepaper from 2014~\cite{Benet14}, is inspired by
various concepts from previous work in networking and file management. It combines a set of
protocols to build a distributed file system on top of a \ac{P2P} network.
For instance, IPFS applies concepts of \ac{ICN}, using uniquely identifiable fingerprints to address
and retrieve files over the \ac{P2P} network rather than location-based references such as \acp{URL}
or IP addresses. However, in contrast to ICN approaches, which primarily use content-centric addressing at the network
layer, the fingerprint-based addressing in IPFS happens at the application layer to ease deployment and guarantee backward compatibility.
The content-centric addressing leverages \emph{Multiformats}, a set of protocols which can generate
\emph{multihashes} to act as a \emph{\acfi{CID}} of an object, and allows nodes to download
different parts of a file from multiple peers in the network instead of a single central server.
As content is not only addressed but also linked via unique hashes, IPFS
supports file versioning through Merkle Trees similar to \emph{Git} as well.



Moreover, IPFS implements its protocols in a stack, which means its components can be extended or
replaced when required. This modularity in the design is
supported by a P2P networking library called \emph{libp2p},
which is part of the IPFS protocol stack: libp2p is a modular, \ac{P2P} networking library for \emph{process
addressing}, with a focus on data transfer processes, supporting implementations for several network and transport-layer protocols. libp2p also integrates protocols for
content and peer routing through a \ac{DHT}, a pubsub protocol, and a content exchange protocol called \emph{Bitswap}.

The modularity and flexibility provided by libp2p
are promising concepts and features that can help with decentralising Internet services
such as cloud storage. For instance, Filecoin (see~\autoref{sub:filecoin}) leverages libp2p to build an incentivised and decentralised storage network on top of IPFS, highlighting the reusability of IPFS' modules.
%

\subsection{Overview: Join, Find, and Transfer}
\textbf{Figure~\ref{fig:content-pub-retr}} outlines the process of publishing and retrieving content within IPFS: A content publisher runs an IPFS node and adds a file to IPFS. To do so, it is first imported into an \emph{\ac{IPLD}}
graph (similar to a Merkle hash tree), which essentially constructs its unique name in form of a content address, the \acf{CID}. In turn,
this CID allows a consumer to find and fetch that file (or parts of it, using the CIDs of the intermediate nodes or leaves of the tree). In the following, the specific steps during the process are discussed in more detail.

\textbf{Joining the IPFS network.} The IPFS network is permissionless, which means that peers can freely join the P2P
network by running a peer node, which is identified by a public/private-key
pair. One of the main content routing systems used by IPFS
is libp2p's
\ac{DHT}. New peers connect to (pre-defined) bootstrap nodes to get initial connections for their routing tables.
The DHT is inspired by Kademlia~\cite{Maymounkov02} as well as Coral DHT~\cite{Freedman03}.

\textbf{Finding Content.} The DHT
allows peers to lookup the unique hash identifier (i.e., its CID) of an object~(\autoref{sub:names}), in order to retrieve a list of peer IDs that hold replicas of the object.
These peer IDs are stored within \emph{provider records} in the DHT and contain information about ways to connect to the peers, for instance their IPv4 or
IPv6 addresses along with the transport protocol and port number. This information is included in
libp2p's \textit{multiaddresses},
which express connectivity primitives for peers to connect to other peers. 

\textbf{Preparing and Transferring Content.}
Any object added to the IPFS network is first converted to an IPLD graph (see~\autoref{sub:names} for details), which means that the object is chunked into smaller pieces. Similar to Merkle hash trees, each chunk has its own
\ac{CID}, which represent the leave nodes and are composed to the overall IPLD graph of the file, resulting in a root \ac{CID} for the file.
The CID of any node in the graph can then be 
announced as part of \textit{want lists}, which represent object requests and are sent to the
different peers that store the chunks to initiate file exchange using IPFS' block exchange protocol
called \emph{Bitswap}~\cite{bitswap}; details on Bitswap are omitted for brevity.
Once the transfer is complete, the integrity of the pieces and the whole file can be verified using
the content's hash fingerprints by reconstructing the IPLD graph. From that point on, the requesting
node caches the retrieved file and becomes a temporary provider for that file
(\autoref{sub:caching}).


\subsection{File Processing and Naming}
\label{sub:names}

\begin{figure}[t]
    \includegraphics[width=\columnwidth]{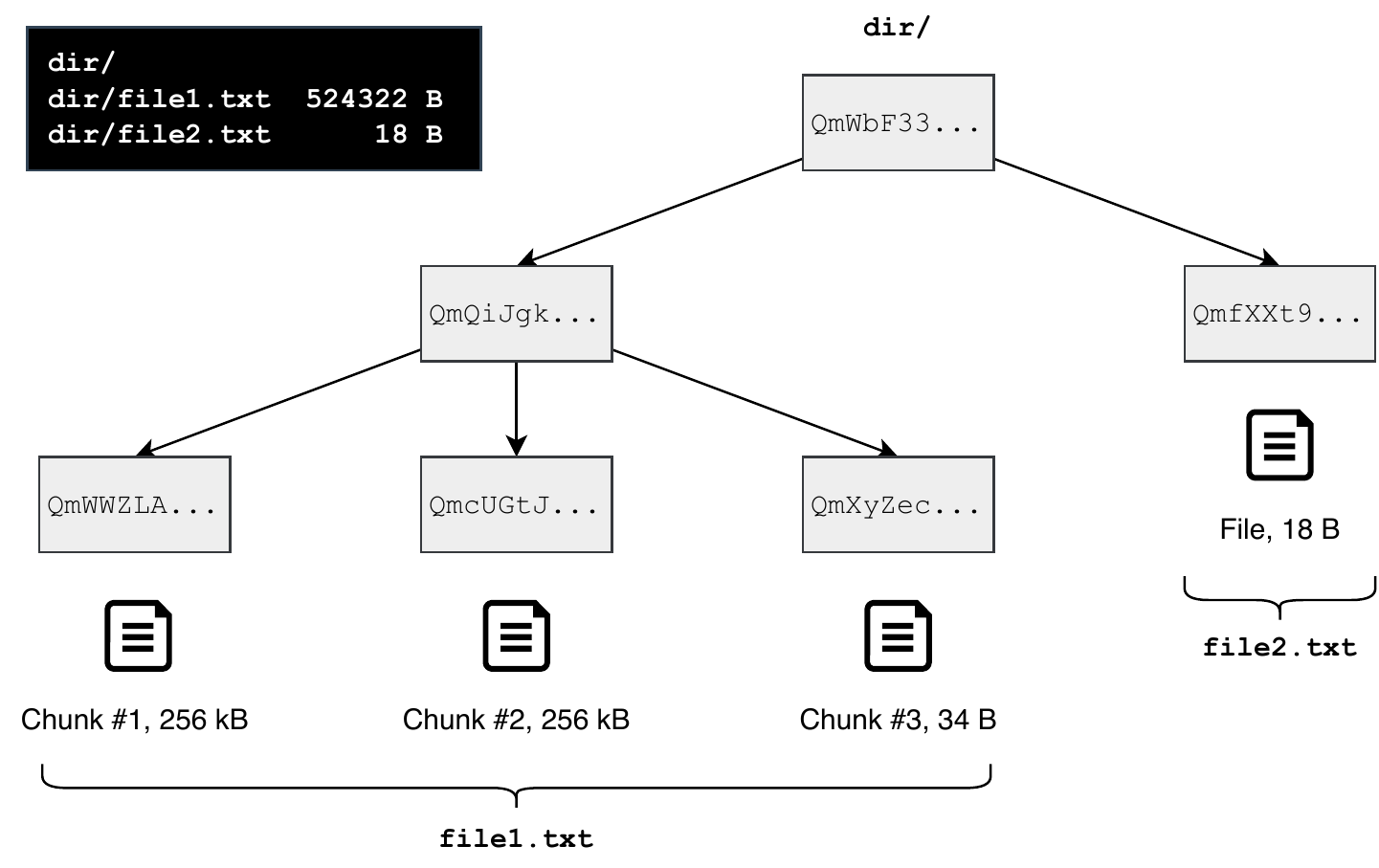}
    \caption{Constructed MerkleDAG for an example directory \texttt{dir/} containing \texttt{file1.txt} and \texttt{file2.txt}, which are chunked into addressable blocks of 256~kB.}
    \label{fig:merkle-dag}
\end{figure}

Any data item added to the network is chunked into blocks of 256~kB (default block size). Each block
is named and addressed individually by a unique IPFS CID. These blocks are arranged as leaf nodes
in a \ac{DAG} to build an IPLD graph, which represents the main data model of IPFS. The root of this hash-linked graph is the unique IPFS
CID for the specific input file (or directory), which allows peers to easily verify the integrity
through the hash-chained DAG (cf. Merkle tree). An example for such a graph and its vertices/edges for a directory with two files is shown in
\textbf{Figure~\ref{fig:merkle-dag}}. Each vertex can be addressed and exchanged individually through its CID, thus, representing the whole directory, the whole files, or individual chunks, respectively.
With this data model, a hash can also to more complex structures
such as subgraphs or other graphs for instance. As such, this naming and addressing structure adopts Linked Data principles which build the foundation of the Semantic Web.
%

The identifier of each block is a self-describing \emph{multihash}, which not only contains
the actual hash digest of the input, but also incorporates the type of hash function that was used,
together with digest length; overall, it prepends both the hash function and the digest length to the digest value. The whole construct is then
encoded in \texttt{base58} to avoid similar looking characters and to obtain alpha-numeric
characters exclusively for unambiguous identifiers. Due to \texttt{SHA-256} being the standard hash
function (assigned to hash type \texttt{0x12}) and digest length being 32 bytes (i.e.
\texttt{0x20}), IPFS fingerprints commonly start with \texttt{Qm...} when encoded in
\texttt{base58}. Files that are hashed with other hash functions can be easily identified and
processed accordingly. Thus, the self-describing multihashes facilitate the implicit replacement of hash functions, which provides backward-compatibility as well as future-proofing.

After processing the file in this manner, a node announces the local storage of each block to the
DHT in order to allow other peers to retrieve the content. In particular, peers that store some
content in the IPFS network produce \emph{provider records}, which they publish on the content routing
system, typically the DHT. Provider records bind the content address to  multiaddresses (e.g., an IP address, plus transport protocol and port combination) and are placed at nodes whose \texttt{peerID} is close to the published object's \ac{CID}, based on the distance between the IDs according to Kademlia, which is calculated through the bitwise exclusive or (XOR) function~\cite{Maymounkov02}. Due to blocks being identified by their multi-hashes, even
smallest modifications will lead to substantially different digest values, which will be recognized during file
integrity verification. 

\subsection{Pathnames and IPNS}
\label{sub:pathnames}
IPFS adopts a pathname scheme with a global namespace similar to the \ac{SFS}~\cite{Mazieres99}.
Once the objects are named, they can be navigated using regular path syntax as commonly employed
in file systems or Web URLs. Towards this end, \texttt{/ipfs/} is added as a prefix to the object
CID to denote that IPFS is used. Following the fingerprint, regular path syntax can be added,
such as \texttt{/ipfs/<CID>/foo/bar.baz} for instance.

As a result of its self-describing, hash-based nature, a CID does not support mutable content.
To avoid having to replace links in cases of file modification and keep the name of content
consistent, \ac{IPFS} supports a naming system called \emph{\ac{IPNS}}. The fingerprint used for IPNS is derived
from a node's public key, which means that each node can only create a single IPNS identifier. While
the identifier itself is static to allow sharing, the CID it refers to can be modified
arbitrarily by the publishing node, which enables mutability for the IPFS object behind the static IPNS identifier; essentially, an IPNS identifier can represent any other arbitrary CID, such as a chunk, file, or data structure. These IPNS identifiers use the \texttt{/ipns/} prefix instead of
the \texttt{/ipfs/} prefix in their pathname to distinguish themselves. 

Alternatively, if clients have access to DNS records and can modify
them, it is also possible to store the IPFS or IPNS identifier in a DNS TXT record and update it
whenever required: A user can publish a DNS TXT record for their domain, e.g. \texttt{domain.name},
containing \texttt{dnslink=/ip[f|n]s/<CID>} to link the domain to a specific IPFS or IPNS
identifier. When accessing a file using \texttt{/ipns/domain.name}, IPNS resolves the respective
\texttt{domain.name} suffix using DNS and replaces it with the stored fingerprint to access the
content.

\subsection{Content Storage and Caching}
\label{sub:caching}
When adding content to the IPFS network as a node, the content is only made available to other peers but
\emph{not} replicated. By default, content is only replicated when it is explicitly requested and
retrieved by another peer, which then caches it locally, with the caching behavior and duration being
configurable. As such, IPFS does not force peers to cache arbitrary content (which they are not interested in
themselves) on behalf of other peers. In other words, peers cache content that they have requested,
following a strict pull model. Among other reasons, this is done to avoid legal implications for the
hosting node. 

Cached content is periodically removed locally by the automatic garbage collection.
In order to become a ``permanent'' provider of some content item, IPFS includes a mechanism called
\textit{pinning}, which allows nodes to mark files as permanent in local storage, i.e., not have them removed by garbage
collection. This means that unpopular content, unless explicitly pinned, will eventually disappear
from the network. Popular content, on the other hand, will be constantly requested and re-cached by peers and will
therefore be disseminated through the network, which additionally improves availability. Due to garbage
collection and churn of nodes, IPFS therefore only provides best-effort storage without any
guarantees for availability.

Overall, as long as peers do not show interest in content a node has published to the IPFS network, the content is \emph{not} disseminated into the network, which is a major difference to most centralized cloud storage solutions. At the same time, this means that responsibility for the availability of the published content is \emph{not} inherently delegated to the network---thus, IPFS does not provide external storage capacity by design, unless content is explicitly requested or pinned by peers.



\subsection{IPFS Public Gateways}
\label{sub:gateway}
In traditional P2P networks, users that are interested in retrieving information from the network or
using its offered services are required to join and participate in the network. In many cases, this
may not be a feasible option, as devices may be resource-constrained, for instance. Thus, peers in
IPFS have the option to act as a gateway for external users who can access the IPFS content using
HTTP(S) instead. It is worth noting that any user in the IPFS network can run a public IPFS gateway, as
long as they have a publicly reachable IP address. 
IPFS gateways serve as a web-server for clients and as a DHT server for the IPFS
network, making access to content seamless.
An example for one of these gateways is
\texttt{ipfs.io}, which can be given either an IPFS or an IPNS CID to request the content. A
regular user would then navigate to \texttt{https://ipfs.io/ip[f|n]s/<CID>} with their
browser to request the object with the respective CID from the \texttt{ipfs.io} gateway. If
not already cached, the gateway retrieves the content from peers in the IPFS network. After retrieving all chunks from the network, the
gateway then serves the file to the requesting user HTTP(S). 

Note that gateways are not essential building blocks by design: they are intended to 
support the network by providing another way of retrieving files from IPFS, in particular to assist
clients which are behind NATs, resource-constrained, or cannot participate as a peer. However,
gateways can also lead to centralisation and dependencies (cf. supernodes in traditional
\ac{P2P} networks), along with free-rider problems (see~\autoref{sub:incentives}).


\section{Properties of IPFS}
\label{sec:properties}

The usage of IPFS for file storage in existing applications can bring a multitude of benefits, such as
built-in file integrity checks, inherent content deduplication, or content-based addressing, which decouples file retrieval from specific
locations. At the same time, it presents special characteristics (such as providing no authorization/access control by default) owing to its decentralised and permissionless nature that need to be considered when integrating
IPFS into a project or application for decentralised storage.
As such, developers using IPFS need to carefully take its
inherent properties, both of the individual protocols within the stack as well as regarding their collective behavior, into account and accommodate them within the remit of their application. We discuss some of the key properties of IPFS and implications in the following.


\subsection{Persistence of Names and File Integrity}
\label{sub:persistence}

Identifying content by a unique multi-hash rather than a location address gives more
flexibility to the network in different ways: resources can be used more
efficiently since duplicate files, and even duplicate blocks of files, are assigned the same
identifier and can therefore be handled, linked, and re-used appropriately to not waste additional
resources. On the other hand, in traditional host-based addressing, duplicate files might end up being stored redundantly under different file names and different location-based identifiers. Note that although centralized cloud providers may also run data deduplication schemes, these schemes are not always applied at the chunk-level and follow different data models~\cite{ShinKH17}.

Explicit content-based
addressing also facilitates on-path and in-network caching, as the integrity of blocks of files can
be verified using the multi-hash. Hence, there is no need to trust third parties to point to, or deliver the
correct file pieces, which circumvents potential centralisation by removing dependency on a single network or (original) content provider.

One property of the persistence of content-based names in IPFS is that content identifiers change when the content itself is updated, e.g., in case of dynamic content. This is in contrast to the current HTTP-based model, where the URL remains when the content it represents changes. Persistence of content names is a desirable property in the design of IPFS, which, thus, creates the need for additional mechanisms to deal with dynamic content. 
The InterPlanetary Naming System \ac{IPNS} (see~\autoref{sub:pathnames}) or libp2p's pubsub protocols, which allow peers to create pubsub channels between each other to dynamically broadcast and listen to events need to be used in order to update content published on IPFS.

There are a few examples of applications on top of IPFS that support dynamic and mutable data:
\emph{Textile} provides a set of developer tools and focuses on data ownership, allowing applications to make use of the data their users provide, e.g., through so called Buckets, which resemble dynamic personalised cloud storage services based on IPFS and libp2p. A similar SDK is \emph{Fission}, which facilitates the publication of frontend apps via built-in backend solutions that handle (user) data management over IPFS.

\subsection{Data Auditability, Censorship Resistance, and Privacy}
One area where IPFS embodies ``privacy by design'' principles more closely than HTTP is in allowing more precise and comprehensive auditability of stored data. For example, in the context of attempting to delete a subject’s personal data after consent has been revoked, a difficulty faced in using HTTP is determining whether all copies of a given piece of data had been deleted from an entity’s servers (as they can be stored under different names). Under IPFS, the persistent nature of content identifiers allows users to know with greater certainty and thoroughness where various files that include associated personal data are stored. Thanks to merkle-linking, one can verify whether an asset is stored in some location by scanning for the asset’s content fingerprint (as well as fingerprints of chunks). However, note that modifications to the data result in different CIDs; thus, modified versions of the data (e.g., with padded chunks) are difficult to detect. Nevertheless, while those content fingerprints themselves are immutable, the actual data can be deleted when needed, for instance to comply with a data subject’s deletion request. This combination --- mutability of data with immutability of certain metadata --- has the potential to provide a more usable basis for applications built on IPFS to comply with both the specific provisions and the broader aims of data protection regulations. 

At the same time, content cannot be explicitly
censored in IPFS, given it operates as a distributed P2P file system with no central indexing entity. Since peers are not organised in a hierarchy, there is no authority node in the
network that can prohibit the storage and propagation of a file, or delete a file from other peers'
storage. Consequently, censorship of unwanted content cannot be technically enforced, which
represents an opportunity for users that are suppressed in their freedom of speech, for instance. Note that censorship of unwanted content within the borders of an oppressive state, for instance, is different to a globally applicable legal request to remove content.
The
censorship resistance that IPFS offers is achieved by replicating content (e.g., snapshots of Wikipedia~\cite{WikipediaIPFS}) among different peers  that
have requested it, which makes it difficult to censor the provider nodes altogether. Moreover, any
public IPFS gateway can also retrieve and deliver the specific content to users, adding further censorship
resistance, especially when hosted by larger \ac{CDN} providers such as Cloudflare; users can simply
use another gateway in case one gateway is being taken down. While this is a fairly simplistic censorship model, note that more sophisticated
censorship approaches also take more advanced information patterns, such as traffic fingerprinting,
into account, which has not been extensively studied for IPFS yet. Nevertheless, due to the
modularity of IPFS in terms of supported/used protocols and possibility to provide content from a
large number of peers, selective censorship is made more difficult over IPFS.


On the other hand, a lookup of a fingerprint to find out which peer stores the file in question can also reveal their IP address, meaning that those peers can still be identified. There are two important points to stress with respect to privacy here:
\begin{enumerate}
    \item As any permissionless, public P2P network, IPFS is a globally distributed, public server for the data published in the network. That said, the primary current use-case of IPFS is providing storage and access to public data, e.g., datasets or websites. Given the modularity and wide range of applications that IPFS envisions to be able to support, IPFS currently does not support privacy at the protocol layer, as applications (such as a public pastebin, for instance) might not require necessarily require privacy. Instead, such privacy designs currently have to be implemented at the application layer instead. Modular approaches to enhance privacy and at the same time support a wide range of applications remains an open research problem at the time of writing.
    \item IPFS peers can control what they share with others in the network. A peer by default announces to the network every CID in its cache, that is, content that they have either published, or have requested and fetched previously. A peer that prefers to keep their request history private can always refrain from re-providing content that they have requested. The peers can still fetch and consume content, as well as keep it in their local cache for later consumption, but they do not serve the content when asked (e.g., through Bitswap) and neither do they let the network (e.g., the DHT) know that they have the particular piece of content locally. Local node configuration
    allows each peer to control what they share with the network.
\end{enumerate}



\subsection{Network Partition Tolerance}
Due to being a decentralised P2P network with no essential central components, IPFS can still
operate in cases of network partitions or in offline scenarios. While some components such as denylists can potentially not be retrieved or updated, partitions do not fully impair the content
publication and retrieval process. Thus, as long as the requested objects and provider records are known and available within the
same subgraph, IPFS is tolerant against partitions and does
not require full Internet connectivity if the providing peers are reachable. Further, private IPFS networks among a set of machines can
be built using \emph{IPFS Cluster},
which allows deployment of IPFS in local networks.

\subsection{Incentives for Participation}
\label{sub:incentives}
IPFS was designed as a permissionless, best-effort, decentralised P2P network and, as such, it does not integrate incentive schemes.
%
%
The operation of an IPFS node incurs costs for infrastructure maintenance in terms of bandwidth,
storage, and power.
After retrieving the desired objects, there is no incentive for a user to keep the node running,
resulting in short sessions and high churn in the network as observed by measurements of the IPFS
network~\cite{Henningsen20}. (Free-riding) consumers may also retrieve the desired objects
conveniently over HTTP through an IPFS gateway instead, which does not require participation in the
network at all. This poses an open research question whether gateways (and similarly supernodes)
arise naturally in a P2P network as a result of trying to support all clients in combination with a
lack of incentives. Nevertheless, incentivising for consistent and continuous participation (and
ultimately a sustainable IPFS network) needs to be considered by the application to avoid
centralisation around gateways and super nodes. \emph{Pinning services}, which are third parties
that pin files to provide improved/guaranteed availability for a monetary return, alleviate the
problem of lacking incentives.
Another way to address the lack of incentives is to provide exclusivity (i.e., content or features that are not available elsewhere); however, this
is difficult to achieve, as user convenience and quality of experience are more unpredictable with the
best-effort storage approach of IPFS in comparison with centralised infrastructures, which may lead
to difficulties in gaining critical mass.
\section{Current Directions and Future Considerations}
\label{sec:future-directions}

The presented properties of IPFS~(\autoref{sec:properties})
highlight various opportunities for
current projects and future considerations. For instance, IPFS is used as off-chain storage for many projects on various blockchains such as Ethereum, where the transactions only contain the immutable IPFS CID, while the data with the respective CID itself is stored on IPFS~\cite{DBLP:conf/ithings/NorvillPSC18}.
The \emph{IPFS Ecosystem Directory}\footnote{\label{fn:ecosystem}\url{https://ecosystem.ipfs.io/}} provides an overview of examples for other existing projects, which range from content delivery over data persistence/governance to social media and e-commerce.
Another project that is closely related and attempts to solve the lack of incentives
along with an improvement of its best-effort content storage is \emph{Filecoin}, briefly discussed in the following.

\subsection{Filecoin}
\label{sub:filecoin}
Filecoin is an incentivised P2P network for the storage and retrieval of
objects that builds on top of IPFS. Its goal is to provide distributed storage which is cheaper than centralised cloud storage
solutions. Filecoin uses the same building blocks~(\autoref{sec:building-blocks}) as IPFS, with the
content addressing via \acp{CID} at its core. Unlike IPFS, which does not replicate content at other
peers unless those peers explicitly request the content, Filecoin provides cryptoeconomic incentives
to its participants: storage and replication of content in the network are rewarded with
cryptocurrency tokens in order to facilitate higher availability, faster retrieval, and to
counteract node churn. In exchange for
a fee, peers can close storage deals between each other to provide persistent storage (cf. service level agreements), which is provable through proof-of-replication and proof-of-spacetime.
Thus, Filecoin improves  IPFS' best-effort
storage and delivery service by providing incentive mechanisms; while aforementioned IPFS pinning services~(\autoref{sub:persistence}) can
also improve IPFS's best-effort approach, Filecoin does not require trust between the parties of a contract
due to its decentralised, blockchain-based foundations. 




% ...

\subsection{Concluding Remarks}
\label{sub:conclusion}

We provided an overview of IPFS and its core features, presenting how its combination of multiple
networking protocols and P2P concepts build a foundation for decentralised cloud storage.
Its building blocks~(\autoref{sec:building-blocks}) enable peer-assisted file distribution and
delivery in order to move away from centralised cloud storages by providing persistence of names,
deduplication, and integrity-checks for files through \acfp{CID}, censorship resistance, network
partition tolerance, and ultimately decentralisation, among other
properties~(\autoref{sec:properties}). 
Previous studies use IPFS for a variety of projects and proposals~\cite{Daniel2022} within the IPFS ecosystem (see Footnote \ref{fn:ecosystem}). Nevertheless, IPFS has yet to overcome
challenges such as access control, participation incentives, or persistent availability and
replication of content. 
Future considerations, such as the integration of IPFS and Filecoin~(\autoref{sec:future-directions}),
aim to overcome some of IPFS' challenges with regard to incentives and content availability. 
Together with the native support of IPFS in the Opera~\cite{ipfs-2021} and Brave~\cite{brave-ipfs} browsers as well as the addition of the \texttt{ipfs://} scheme in Mozilla Firefox~\cite{FirefoxSupport}, these are important steps in stimulating the growth
of the network and moving towards a more decentralised Internet in the future.
The performance of these decentralised solutions is a very timely research topic, which we encourage the community to undertake. As such, future work on distributed storage in general should
consider both opportunities and challenges of IPFS in order to develop suitable decentralised
storage systems for the Future Internet and its applications. Based on the discussion presented in this paper in particular, we identify several broad,
open research questions with respect to IPFS and distributed storage in general: How does P2P-based content
storage and retrieval compare to traditional cloud storage or \acp{CDN} technologies? 
How can availability, retrieval, and delivery of content be improved? 
How can we achieve full user anonymity when fetching and retrieving content in permissionless P2P networks? 
How can the use and adoption of such
decentralised technologies be incentivised?




\bibliographystyle{IEEEtran}
\bibliography{ipfs,ipfs-review-dblp,ipfs-review-ieee-acm}

\begin{IEEEbiography}{Trinh Viet Doan,}{\,}is a PhD student at the Technical University of Munich, Germany. His research interests include Internet measurements as well as consolidation and decentralization of Internet architectures and infrastructures. He received his Master's degree from Technical University of Munich in 2017. Contact him at doan@in.tum.de.
\end{IEEEbiography}

\begin{IEEEbiography}{Dr. Yiannis Psaras,}{\,}is a research scientist in the Resilient Networks Lab at Protocol Labs. His research interests include Information- or Content-Centric Networks, as well as resource management techniques for current and future networking architectures with particular focus on routing, caching, and congestion control. He received his PhD degree from Democritus University of Thrace in 2008. Contact him at \mbox{yiannis@protocol.ai}.
\end{IEEEbiography}

\begin{IEEEbiography}{Prof. Dr.-Ing. Jörg Ott,}{\,}has been the Chair of
Connected Mobility, Faculty of Informatics, Technical University of Munich,
since August 2015. His research interests are in network architectures, (transport) protocols, and algorithms for connecting mobile nodes to the Internet and to each other. He received his PhD degree from TU Berlin in 1997. Contact him at ott@in.tum.de.
\end{IEEEbiography}

\begin{IEEEbiography}{Dr. Vaibhav Bajpai,}{\,}is an independent research group leader at CISPA Helmholtz Center for Information Security, Hannover. His current research focuses on improving Internet operations (e.g., performance, security, and privacy) using data-intensive methods and by building real-world systems and models. He received his PhD degree from Jacobs University Bremen in 2016. Contact him at bajpai@cispa.de.
\end{IEEEbiography}

\vfill

\end{document}